\documentclass[pre,twocolumn,superscriptaddress,showpacs,aps,floatfix]{revtex4}
\usepackage{graphicx}

\begin{document}

\title{Enhanced Transmission and Giant Faraday Effect in Magnetic
Metal-Dielectric Photonic Structures}
\author{Kyle Smith}
\affiliation{Department of Physics and Astronomy, University of Texas at San Antonio, San
Antonio, TX 78249, USA}
\author{Turhan Carroll}
\affiliation{Air Force Research Laboratory, Sensors Directorate, Wright Patterson AFB, OH
45433, USA}
\author{Joshua D. Bodyfelt}
\affiliation{ElectroScience Laboratory \& Department of Electrical Engineering, The Ohio
State University, Columbus, OH 43212, USA}
\author{I. Vitebskiy}
\affiliation{Air Force Research Laboratory, Sensors Directorate, Wright Patterson AFB, OH
45433, USA}
\author{A. A. Chabanov}
\affiliation{Department of Physics and Astronomy, University of Texas at San Antonio, San
Antonio, TX 78249, USA}
\affiliation{Air Force Research Laboratory, Sensors Directorate, Wright Patterson AFB, OH
45433, USA}
\date{\today}

\begin{abstract}
Due to their large electrical conductivity, stand-alone metallic films are
highly reflective at microwave frequencies. For this reason, it is nearly
impossible to observe Faraday rotation in ferromagnetic metal layers, even
in films just tens of nanometers thick. Here, we show using numerical
simulations that a stack of cobalt nano-layers interlaced between dielectric
layers can become highly transmissive and display a large Faraday rotation and
extreme directionality. A 45-degree Faraday rotation commonly used in
microwave isolators can be achieved with ferromagnetic metallic layers as
thin as tens of nanometers.
\end{abstract}

\maketitle

%78.20.Ls - Magneto-optical effects
%41.20.Jb - Electromagnetic wave propagation; radiowave propagation
%78.67.Pt - Multilayers; superlattices; photonic structures; metamaterials
%78.66.Bz - Metals and metallic alloys
%%
\section{Introduction}
Magnetic materials are of great importance to microwave (MW) engineering and
optics due to their nonreciprocal properties, such as the Faraday and Kerr
effects (see, e.g., \cite{Zvezdin,Pozar} and references therein). These
properties are utilized in various nonreciprocal devices including
isolators, circulators, phase shifters, etc. The functionalities of these
devices stem from the intrinsic gyrotropy of their magnetic components, such
as ferrites and other magnetically polarized materials, in which left and
right circularly polarized waves experience nonreciprocal differential phase
shifts and/or attenuation. When the magnetic material is placed in a
resonator or introduced in a photonic structure, the nonreciprocal response
can be significantly enhanced \cite{Inoue1998,Sakaguchi1999}. A noticeable
enhancement of the magneto-optical Faraday effect was observed in magnetic
thin-film layers sandwiched between Bragg reflectors \cite%
{Inoue1999,Shimizu2000,Hamon2006}, periodic multilayers \cite%
{Grishin2004,Shuvaev2011} and other magnetophotonic structures supporting
slow and localized modes \cite{Goto2008}. In all those cases, however, the
Faraday rotation enhancement was accompanied by a significant decrease in
the optical transmittance \cite%
{Inoue1999,Shimizu2000,Hamon2006,Grishin2004,Shuvaev2011,Goto2008}. Although
the transmittance can be somewhat improved by minimizing the reflectance of
the photonic structures \cite{Sakaguchi1999,Steel2000a,Steel2000b}, the
output can still be severely affected by absorption losses \cite%
{Kato2003,Atkinson2006}. This limitation stems from the fact that at optical
frequencies, both the nonreciprocal response and photon absorption are
governed by the permittivity tensor of the magneto-optical material \cite%
{Zvezdin}. Thus, any enhancement of the Faraday effect is inevitably
accompanied by similar enhancement of absorption losses.

At MW frequencies, the nonreciprocal response is usually associated with the
magnetic permeability tensor of the gyrotropic magnetic material, whereas
the absorption is often caused by the electric conductivity of the magnetic
material. In other words, the magnetic component of the electromagnetic wave
is responsible for the nonreciprocal light-matter interactions, while the
electric component -- for the absorption losses. In this case, a
properly designed magneto-photonic structure can enhance magnetic Faraday
effect, while drastically reducing the absorption losses \cite{Vitebsky2008}. 
Indeed, in the vicinity of a resonance of magnetophotonic structure, the nodes
of the electric field component usually coincide with the antinodes of the
magnetic field component, as shown in Fig. 1.
%%%%%%%%%%%%%%%%%%%%%%%%%%%%%%%%%%%%%%%%%%%%%%%%%%%%%%%%
\begin{figure}[hbt]
\includegraphics[width=0.8\columnwidth, keepaspectratio]{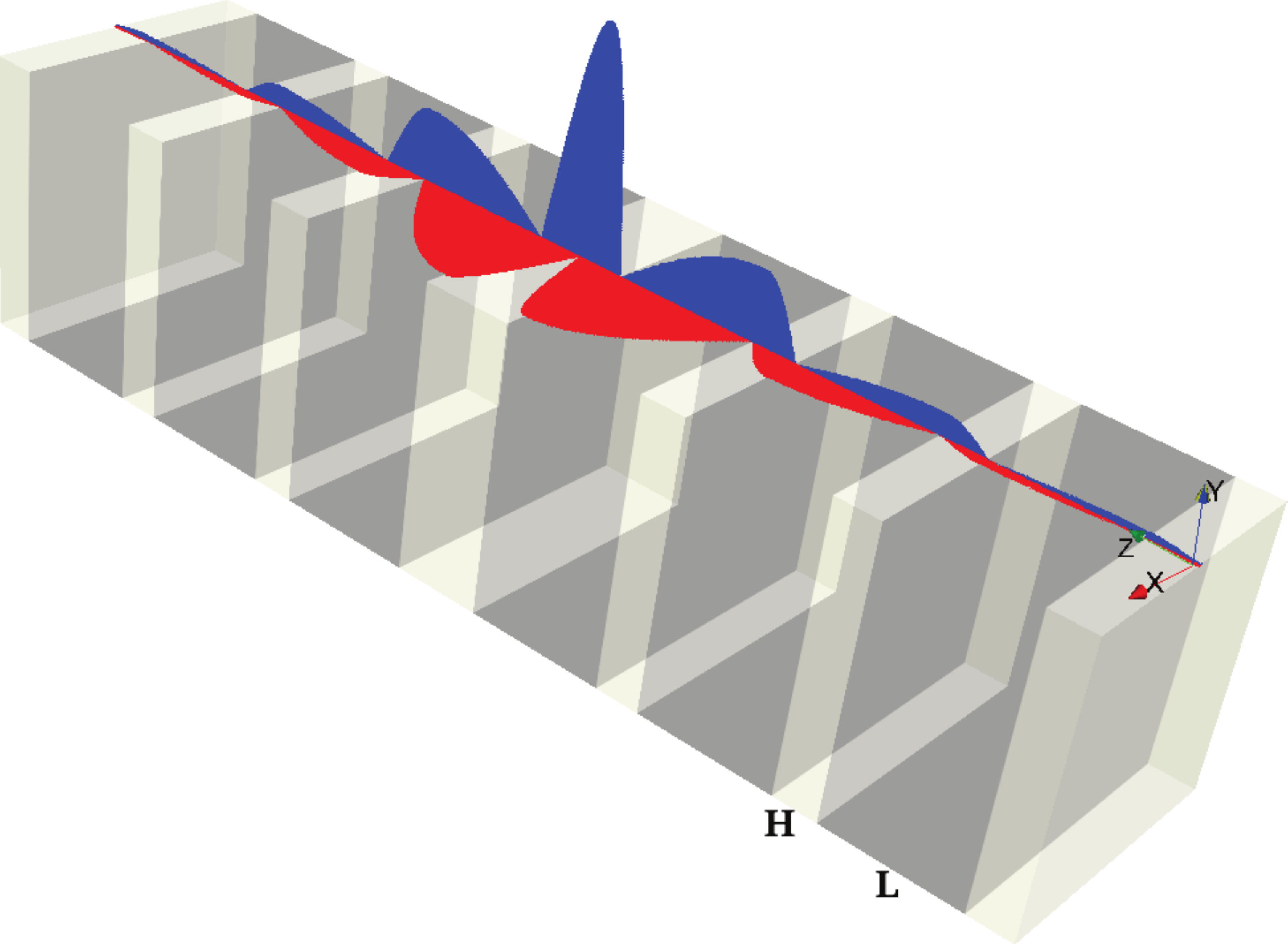}
\caption{Schematic representation of the quarter-wave dielectric
layer stack $3:3$ with a central half-wave defect and spatial distributions
of the electric field $e_x(z)$ (red) and magnetic field $h_y(z)$ (blue)
components of the defect localized mode. The electric field is seen to have
a node at the middle of the defect, whereas the magnetic field has a maximum
at the same location. $H$ and $L$ indicate high (alumina) and low (air)
refractive index constitutive layers of the structure.}
\end{figure}
%%%%%%%%%%%%%%%%%%%%%%%%%%%%%%%%%%%%%%%%%%%%%%%%%%%%%%%%
Therefore, when a conducting
magnetic layer is positioned at a node of the electric field, the Ohmic
losses will be suppressed dramatically, while the nonreciprocal (magnetic)
light-matter interactions will be enhanced. The nodes and antinodes of the
electric and magnetic components of electromagnetic wave are only
well-defined at high-Q resonances or under slow-wave conditions. In
addition, the magnetic layer displaying a significant electric conductivity
should be thin enough to fit in the electric field node. Otherwise, the
Ohmic losses will be prohibitively large, and the Faraday rotation
enhancement will also be compromised. In order to use a very thin magnetic
layer and to produce a sufficiently large Faraday rotation, the magnetic
material should have strong enough circular birefringence.

In this work we use numerical simulations to demonstrate that ferromagnetic
metals, such as cobalt, meet the above conditions and can be utilized in
magneto-photonic structures to produce large Faraday rotation with
negligible losses. This is in spite the fact that stand-alone metallic
layers are almost completely reflective at MW frequencies -- even when the layers are so
thin (a few tens of nanometers) that they do not produce any measurable
Faraday rotation. For this reason, it is nearly impossible to observe
Faraday rotation in ferromagnetic metals. Here, we demonstrate that when one
or a few cobalt nano-layers are incorporated into a properly designed stack
of dielectric layers, the entire layered structure not only can be highly
transmissive, but it can also produce a large Faraday rotation. In addition,
a properly designed metallo-dielectric, magneto-photonic structure can
feature extreme directionality, transmitting the incident radiation
propagating only in a single direction, e.g., the forward direction along
the $z$ axis (the $+z$ direction). The radiation incident from all other
dirctions, including $-z$, will not pass through the structure. This unique property 
owns its very existence to the high electric conductivity of
the magnetic material. In other words, the affect of electric conductivity
of the magnetic material is not just what we want to suppress. Quite the
opposite, the conductivity, along with the magnetic Faraday effect, are
essential in providing the extreme transmission directionality. Finally, we
briefly discuss coupled-resonance structures in which high transmission and
large Faraday rotation can be achieved over a finite frequency range.
\section{Magnetic Metal-Dielectric Photonic Structures}
MP structures that are studied here are constructed as the following. We
consider 1D MW photonic crystals consisting of layers of alumina ($H$) and
air ($L$) of dielectric constants $\epsilon_H=10$ and $\epsilon_L=1$,
respectively, and magnetic permeability $\mu=1$. Since alumina and air have
negligible absorbing at MW frequencies, we take them to be lossless in our
numerical simulations. The dielectric layers have quarter-wave thicknesses $%
d_j=\lambda_0/4\sqrt{\epsilon_j}$, for $j=H,L$, at the wavelength $%
\lambda_0=4$ cm corresponding to the midgap frequency $f_0=7.5$ GHz of the
quarter-wave layer stack. Starting with a periodic stack $HLH\ldots L$ of $%
M=2N+1$ unit cells ($N$ is a positive integer), we add layer $H$ at the end
of the stack and remove the middle layer $L$, to construct a symmetric layer
stack with a central half-wave defect. The resulting structure is designated
$N:N$ (Fig. 1 depicts $3:3$ structure). 

The half-wave defect introduces a
pair of polarization-degenerate localized states at the midgap frequency $%
f_0 $. The spatial profiles of the electric and magnetic field amplitudes of
the localized mode of the $3:3$ structure are shown in Fig. 1. Whereas the
electric field is seen to have a node at the middle of the defect, the
magnetic field is sharply peaked at the same position. The MP structure $NCN$
is then constructed by inserting a cobalt ($C$) layer in the node of the
electric field at the middle of the defect. We also consider periodic
arrangements with multiple defects ($N:M:N$, $N:M:M:N$, etc.) and multiple
cobalt layers ($NCMCN$, $NCMCMCN$, etc.)

The MW permittivity of cobalt is large and almost purely imaginary, $%
\epsilon_C\approx i4\pi\sigma_C/\omega$, where $\omega$ is the angular
frequency, and the electric conductivity of cobalt is $\sigma_C=1.44\times
10^{17}$ s$^{-1}$ \cite{CRC}. Stand-alone metallic layers are highly
reflective to MW radiation even at thicknesses far less than the skin depth,
due to multiple reflections in the metallic layer \cite{Kaplan1964}. The MW
transmittance of a cobalt layer of thickness $d_C$ at frequencies not too
close to the ferromagnetic resonance can be approximated by $T\approx(1+2\pi
d_C\sigma_C/c)^{-2}$, where $c$ is the speed of light. For example, for $%
d_C=33$ nm, $T\sim 10^{-4}$.

The MW permeability of cobalt is described by the Polder permeability tensor
\cite{Polder1949}. We assume that a static uniform magnetic field is applied
perpendicular to the cobalt layer along the wave propagation direction ($z$%
-axis in Fig. 1) and that the internal static field, $H_i$, saturates the
cobalt layer whose magnetization $M$ is oriented along the $z$-axis. In the
presence of the MW field, $\mathbf{h}$ ($h\ll H_i$), which is applied
perpendicular to the $z$-axis, the ac component of the internal flux
density, $\mathbf{b}$, is given by $\mathbf{b}=\hat\mu\mathbf{h}$, where the
permeability tensor $\hat\mu$ is
\begin{equation}
\hat{\mu}=\left|
\begin{array}{ccc}
\mu & i\kappa & 0 \\
-i\kappa & \mu & 0 \\
0 & 0 & 1%
\end{array}
\right| ,
\end{equation}
with the elements \cite{Gurevich}
\begin{equation}
\mu=1+\frac{\omega_{M}(\omega_{H}+i\alpha\omega)}{(\omega_{H}+i\alpha%
\omega)^{2}-\omega^{2}}\,,\,\,\,\, \kappa=\frac{\omega_{M}\omega}{%
(\omega_{H}+i\alpha\omega)^{2}-\omega^{2}}\,,
\end{equation}
where $\omega_{H}=\gamma H_i$, $\omega_{M}=\gamma4\pi M$, $\gamma/2\pi=2.8$
GHz/kOe is the gyromagnetic ratio, and $\alpha$ is the dissipation
parameter. Here we use the cobalt parameters \cite{Frait1965}, $4\pi
M=17\,900$ G and $\alpha=0.027$.

The nonreciprocity of cobalt is determined via the off-diagonal elements of
the permeability tensor. One can show that for circularly polarized waves
the permeability tensor becomes diagonal with the effective permeabilities
\begin{equation}
\mu_{\pm}=\mu\pm\kappa=1+\frac{\omega_{M}}{\omega_{H}+i\alpha\omega\mp\omega}%
\,,
\end{equation}
where the resonant behavior occurs only for the right (+) circularly
polarized wave. For $\alpha\ll 1$, the imaginary part of $\mu_+$, which is
responsible for magnetic losses, is important only in the vicinity near the
ferromagnetic resonance, $\omega^{2}_H=(1+\alpha^2)\omega^{2}$, where it
develops a steep maximum \cite{Hogan1953}. On the other hand, the real part
of $\kappa$, $\kappa^{\prime}=\Re k$, which is responsible for magnetic
circular birefringence, remains considerable even far from resonance where
it can be approximated as $\kappa^{\prime}\approx
\omega_{M}\omega/(\omega_{H}^{2}-\omega^{2})$.

We consider a linearly polarized MW field normally incident on MP
structures. Designating the incident, reflected and transmitted electric
field components by $E_i$, $E_\rho$ and $E_\tau$, respectively, we define
the complex reflection and transmission coefficients of the MP structure, $%
\rho=|\rho|\exp(i\varphi)$ and $\tau=|\tau|\exp(i\phi)$, by $E_\rho=\rho E_i$
and $E_\tau=\tau E_i$. The reflectance $R$ and transmittance $T$ are given
by $R=|\rho|^2$ and $T=|\tau|^2$, and the absorptance $A$ is $A=1-R-T$.
Since the linearly polarized incident wave can be considered as composed of
two circularly polarized components of equal amplitude, the reflected and
transmitted waves are in general elliptically polarized due to the
difference in the permeabilities $\mu_{+}$ and $\mu_{-}$ of Eq. (3). For the
transmitted wave, the Faraday rotation of the plane of polarization of the
wave relative to the incident polarization is $\theta_{FR}=(\phi_{+}-%
\phi_{-})/2$, and the Faraday ellipticity is $e_F=(|\tau_{+}|-|\tau_{-}|)/(|%
\tau_{+}|+|\tau_{-}|)$ \cite{Palik1970}. The transfer-matrix method is used
to calculate the variation of $\rho$ and $\tau$ with wave frequency $f$,
cobalt thickness $d_C$ and magnetic field $H_i$.
\section{Results and Discussion}
%%%%%%%%%%%%%%%%%%%%%%%%%%%%%%%%%%%%%%%%%%%%%%%%%%%%%%%%
\begin{figure}[htb]
\includegraphics[width=0.8\columnwidth, keepaspectratio]{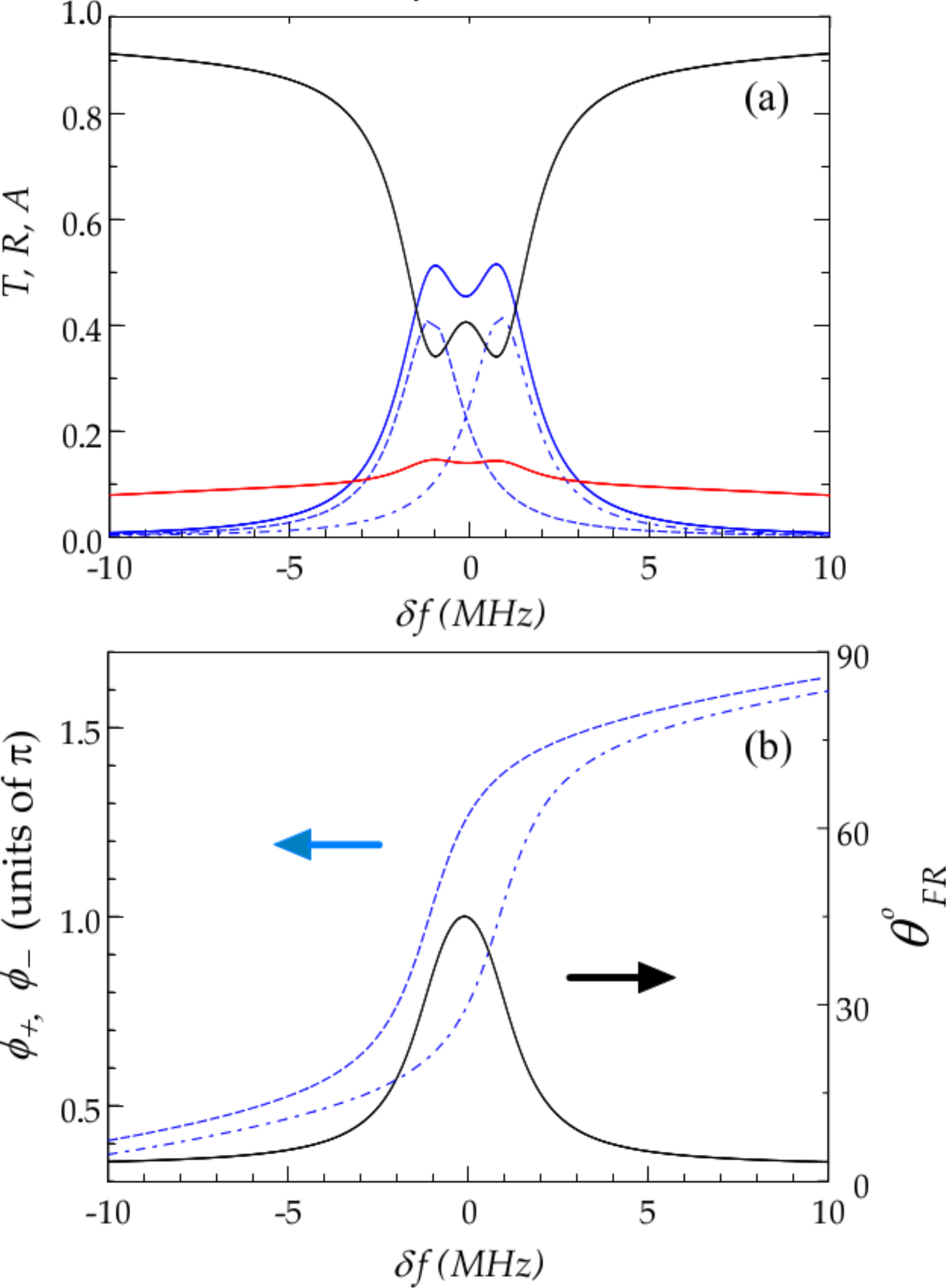}
\caption{Frequency response of $3C3$ structure ($3:3$ structure
with a cobalt layer added into the defect) to a linearly-polarized, normally
incident MW field. Upper panel: transmittance, $T$, (blue), reflectance, $R$%
, (black) and absorptance, $A$, (red) are plotted versus the detuning $%
\delta f$ from the midgap frequency $f_0=7.5$ GHz; left and right circularly
polarized transmission contributions, $|\tau_+|^2$ (dashed line) and $%
|\tau_-|^2$ (broken line). Lower panel: phase spectra of the circularly
polarized components, $\phi_{+}$ (dashed) and $\phi_{-}$ (broken), and
Faraday rotation, $\theta_{FR}=(\phi_+-\phi_-)/2$. The $3C3$ structure was
optimized for a pure $45^0$ Faraday rotation with $d_C=180$ nm and $H_i=0.1$
kOe.}
\end{figure}
%%%%%%%%%%%%%%%%%%%%%%%%%%%%%%%%%%%%%%%%%%%%%%%%%%%%%%%%
\subsection{Normal incidence}
The frequency response of the $3C3$ structure in the vicinity of the
transmission resonance as a function of detuning $\delta f=f-f_0$ is
presented in Fig. 2. 

The transmission of linearly polarized incident wave
(solid blue line) can be decomposed into the sum of two components, $%
|\tau_{+}|^2$ and $|\tau_{-}|^2$, shown in Fig. 2(a) by dotted and broken
lines, respectively. The splitting of the transmission resonance into two
peaks is caused by the difference in the magnetic permeability for the left
and right circularly polarized waves propagating in the direction of $M$.
The amount of the splitting depends on the cobalt thickness $d_C$ and the
magnetic field $H_i$. It increases with increasing $d_C$ and as $H_i$
approaches the resonance field, $2\pi f_0/\gamma$. In the vicinity of the
resonance peak, the transmission phase ($\phi_{+}$ and $\phi_{-}$, shown in
Fig. 2(b) by the dotted and broken lines, respectively) increases by $\pi$,
so that the Faraday rotation $\theta_{FR}$ (shown by solid black line) has a
maximum at $\delta f=0$. Note that the pure Faraday rotation of linearly
polarized wave, i.e., with no wave ellipticity, occurs when $%
|\tau_+|=|\tau_-|$, and thus falls between the resonance peaks. The $3C3$
structure can be optimized for the pure 45$^0$ Faraday rotation by adjusting
the separation of the resonance peaks with a proper combination of $d_C$ and
$H_i$. A $\pi/2$-phase difference occurs at the midpoint between the two
resonances at which $|\tau_+|=|\tau_-|$ when the resonance separation is
equal to the FWHM of the resonance. In Fig. 2, the $3C3$ structure is
optimized for a pure 45$^0$ Faraday rotation with $d_C=180$ nm and $H_i=0.1$
kOe. The corresponding transmittance $T_{45^0}$ is 0.46, which is only
slightly below $T_{45^0}=0.5$ for the case of no absorption. The absorptance
and reflectance are shown in Fig. 2(a).

%%%%%%%%%%%%%%%%%%%%%%%%%%%%%%%%%%%%%%%%%%%%%%%%%%%%%%%%
\begin{figure}[hbt]
\includegraphics[width=0.75\columnwidth, keepaspectratio]{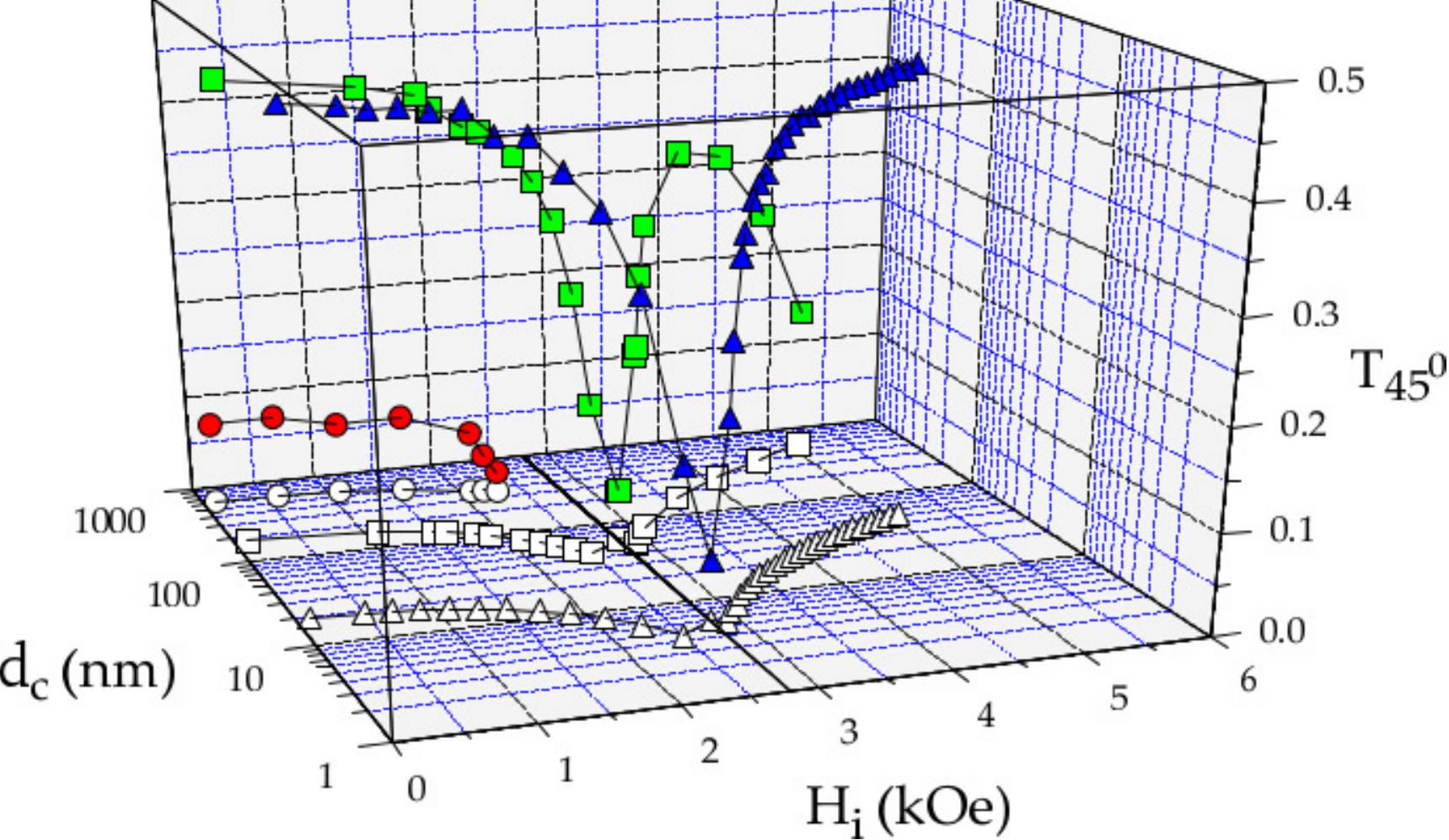}
\caption{$NCN$ structure parameters, cobalt film thickness $d_C$
and internal static field $H_i$ (open symbols), optimized for a pure $45^0$
Faraday rotation and associated transmittance $T_{45^0}$ (filled symbols)
for $N=2$ (circle), 3 (square), and 4 (triangle).}
\end{figure}
%%%%%%%%%%%%%%%%%%%%%%%%%%%%%%%%%%%%%%%%%%%%%%%%%%%%%%%%
For $NCN$ structures, there is a range of combinations of $d_C$ and $H_i$
for the pure 45$^0$ Faraday rotation. A few such combinations with
corresponding values of $T_{45^0}$ are shown in Fig. 3 for $N=2$ (circles), $%
N=3$ (squares), and $N=4$ (triangles). Except for $N=2$, high values of $%
T_{45^0}$ occur both above and below the resonant field, $2\pi
f_0/\gamma=2.68$ kOe, at which $T_{45^0}$ vanishes due to magnetic circular
dichroism. Note that in Fig. 3 suitable values of the cobalt thickness $d_C$
differ by orders of magnitude for different $N$, falling to tens of
nanometers for $N=4$.

%%%%%%%%%%%%%%%%%%%%%%%%%%%%%%%%%%%%%%%%%%%%%%%%%%%%%%%%
\begin{figure}[h]
\includegraphics[width=0.8\columnwidth, keepaspectratio]{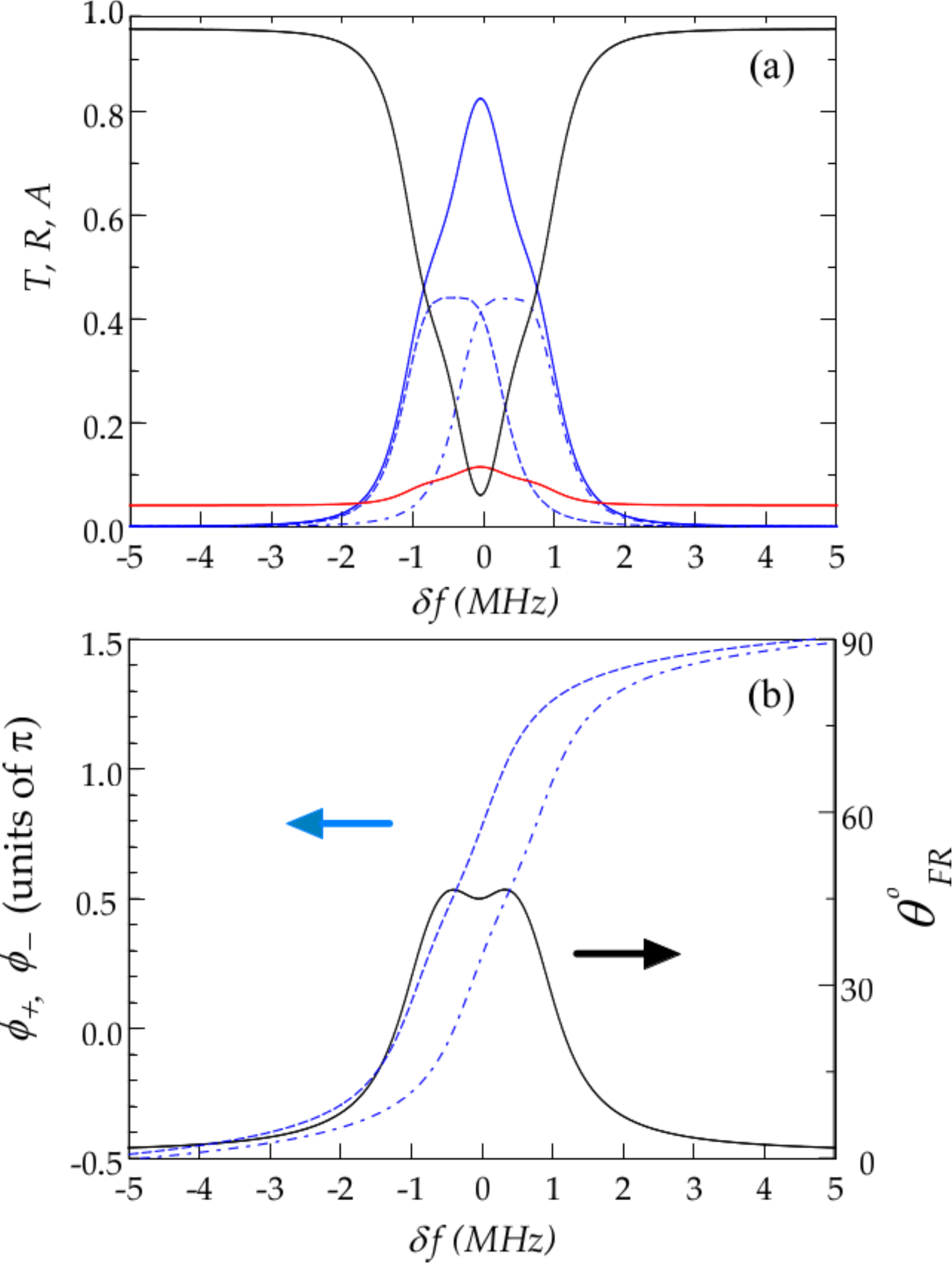}
\caption{Same as in Fig. 2, but for $3C7C3$ structure ($3:7:3$
structure with identical cobalt layers added into the defects). Note a
decreased range of the frequency axis compared to Fig. 2. The $3C7C3$
structure was optimized for a pure $45^0$ Faraday rotation with $d_C=70$ nm
and $H_i=0.1$ kOe.}
\end{figure}
%%%%%%%%%%%%%%%%%%%%%%%%%%%%%%%%%%%%%%%%%%%%%%%%%%%%%%%%
The periodic arrangements with multiple defects, $N:M:N$, $N:M:M:N$, etc.,
can be considered as a superlattice of $N:N$ structures coupled via
phase-matching air layers, or coupled-resonance structures (CRS) (cf. \cite%
{Yariv1999}). As a result of the coupling, the eigenmodes of individual $N:N$
structures split into a miniband of polarization-degenerate localized states
centered at the midgap frequency $f_0$ of the $N:N$ structure. The miniband
width depends on the coupling strength, and thus on $N$. One of the most
important properties of CRS is that the dispersion relation and transmission
spectrum for a moderate number of unit cells of the superlattice can be
optimized to closely approximate that of the infinite CRS \cite%
{Ye2004,Ghulinyan2006}. As a result, the transmission phase via a
finite-size CRS exhibits a smooth increase through a finite and high
transmission miniband summing up to a total phase shift equal to the number
of half-wave defects in the structure multiplied by $\pi$ \cite%
{Ghulinyan2006}. When magnetic layers are introduced into the finite-size
CRS, it is possible to achieve a uniform Faraday rotation in a finite and
high transmission miniband.

The frequency response of the $3C7C3$ structure (2-unit cell CRS) is
presented in Fig. 4. As in the $3C3$ structure, the cobalt layers lift the
polarization degeneracy of localized modes, resulting in splitting of
transmission resonances for $|\tau _{+}|^{2}$ and $|\tau _{-}|^{2}$ (Fig.
4a). In contrast to the $3C3$ structure, however, each of the two resonances
represents a band of two overlapping localized modes. The transmission
phases $\phi _{+}$ and $\phi _{-}$ increase by $2\pi $ through the
resonances (Fig. 4b), and the pure $45^{0}$ Faraday rotation can be achieved
with a smaller resonance separation (note the different scales on the
frequency axes in Figs. 2 and 4). This, in turn, leads to a significantly
higher transmittance, $T_{45^{0}}=0.82$, as compared to the $3C3$ structure.
The results of Fig. 4 are obtained with $d_{C}=70$ nm and $H_{i}=0.1$ kOe.
We note that a wider and higher transmission miniband of the pure $45^{0}$
Faraday rotation can be obtained with a larger number of unit cells of the
CRS.

Furthermore, when the mgnetophotonic structure $3C3$ or $3C7C3$ is placed between linear
polarizers with a $45^{0}$ misalignment angle between their acceptance
planes, the entire structure acts as a MW isolator, as illustrated in Fig. 5.
%%%%%%%%%%%%%%%%%%%%%%%%%%%%%%%%%%%%%%%%%%%%%%%%%%%%%%%%
\begin{figure}[htb]
\includegraphics[width=0.8\columnwidth, keepaspectratio]{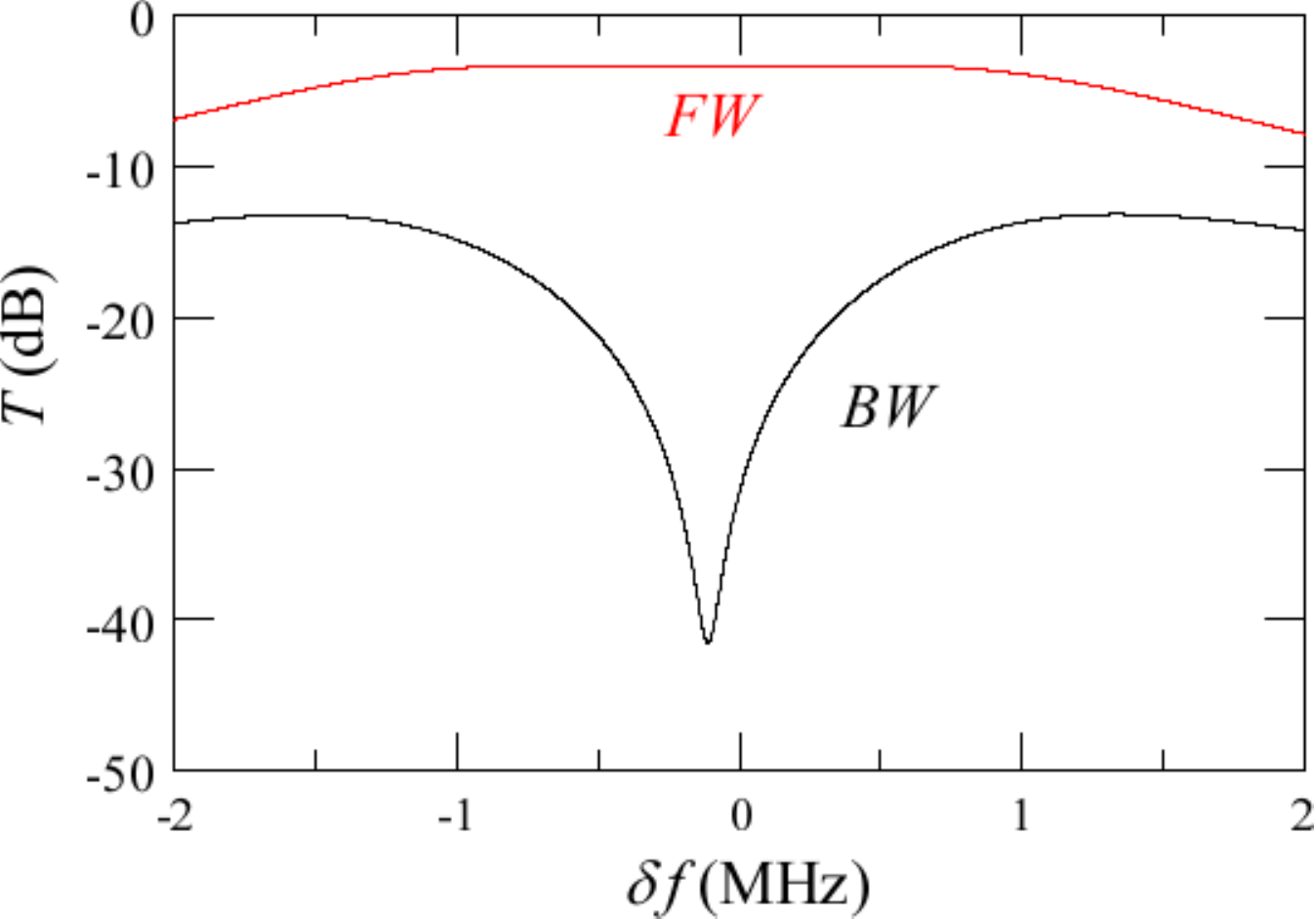}
\caption{Forward (FW) and backward (BW) transmission via a $3C3$
structure sandwiched between linear polarizers oriented so that their
acceptance planes make an angle of $45^0$ with each other. A dip in the
backward transmission (i.e., isolation), occurs for a $45^0$ Faraday
rotation induced by the $3C3$ structure and is limited by a small
ellipticity of the transmitted wave. The $3C3$ structure is optimized for a $%
45^0$ Faraday rotation with $d_C=180$ nm and $H_i=0.1$ kOe.}
\end{figure}
%%%%%%%%%%%%%%%%%%%%%%%%%%%%%%%%%%%%%%%%%%%%%%%%%%%%%%%%
%%
\subsection{Oblique incidence }
%%
%%%%%%%%%%%%%%%%%%%%%%%%%%%%%%%%%%%%%%%%%%%%%%%%%%%%%%%%
\begin{figure}[h]
\includegraphics[width=0.8\columnwidth, keepaspectratio]{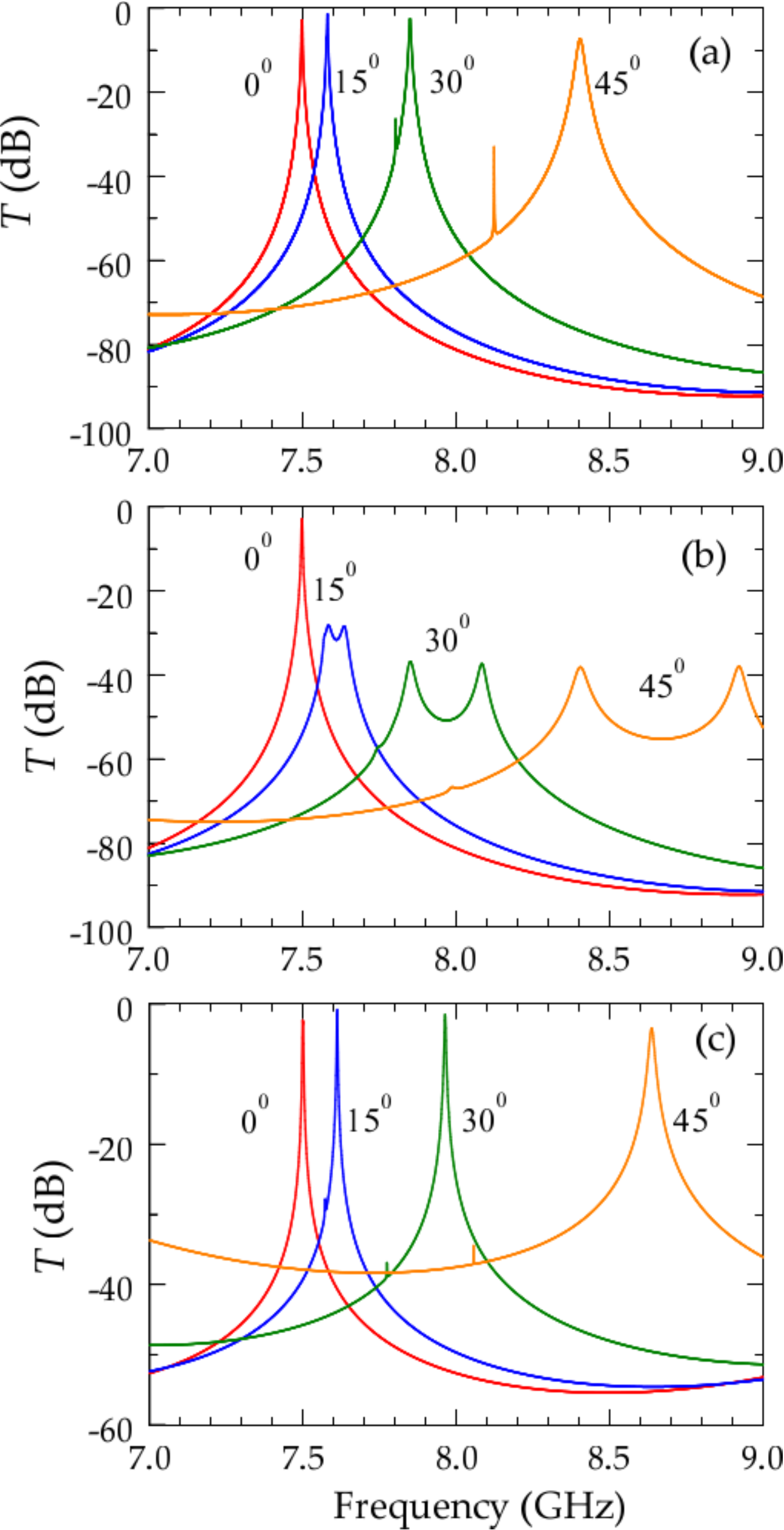}
\caption{Transmission spectra of the $3C3$ structure with symmetric (a) and asymmetric (b) defect at oblique incidence. In the latter case, as the angle of incidence increases, the
transmission spectra are increasingly blue shifted and strongly reduced due to reflection from the metallic cobalt layer, as compared to that for normal incidence. (c), when the
cobalt layer is replaced by a non-conductive magnetic layer of the same thickness, the magneto-photonic structure remains transmissive.}
\end{figure}
%%%%%%%%%%%%%%%%%%%%%%%%%%%%%%%%%%%%%%%%%%%%%%%%%%%%%%%%
Let us now briefly discuss the case of oblique incidence. At
oblique incidence, the effect of induced transparency does not go away, but
the respective resonance frequency increases with the incident angle,
as shown in Fig. 6(a). Qualitatively, this behavior is the same for both TE
and TM polarizations. If we keep the frequency constant, the 1D magnetic 
metal-dielectric photonic structure will act as a collimator, transmitting 
EM radiation only along a single direction. This transmission direction, though, 
will depend on frequency.

At the resonant frequency, the position of metallic nano-layer must coincide with 
the node of the electric field. In all cases involving the symmetric structure of Fig. 1, 
the resonance conditions automatically imply that the introduced metallic nano-layer is indeed located at the
electric field node. If we modify the defect by making it
asymmetric, the induced transparency vanishes for all
frequencies and all incident directions. We can modify the defect layers, however, so that
the induced transparency only persists at normal incidence, as shown in Fig.
6(b). In such a case, the layered structure becomes extremely directional
and only transmits the incident radiation at normal incidence and within a
narrow frequency band. When such a structure is placed between linear polarizers with a $45^{0}$ misalignment between each other, 
it only passes the radiation along the $+z$ direction. We emphasize that such an extreme
directionality of the asymmetric layered structure is caused by high
electric conductivity of the metallic nano-layer. Indeed, when the
cobalt nano-layer is replaced by a non-conducting magnetic material of the same magnetic properties (see Fig. 6c), 
the effect of extreme directionality of Fig. 6b goes away.
\section{Conclusion}
In this paper, we analyzed the scatterring problem of a plane
electromagnetic wave normally incident on a layered structure containing
ferromagnetic metallic nano-layers. The  location of the metallic
nano-layers inside the resonant cavity coincides with the electric field
node, which suppresses the Ohmic losses by four to six orders of magnitude.
At the same time, the magnetic Faraday rotation is enhanced because the
ferromagnetic nano-layers are located at the antinodes of the oscillating
magnetic field. As a result, the layered structure becomes nearly perfectly
transmissive, while producing a strong, 45$^{0}$ Faraday rotation. By
contrast, depending on its thickness, a stand-alone ferromagnetic metallic
layer would be either totally reflective, or would not produce any
measurable Faraday effect at all. In addition, a combination of high
electric conductivity  and magnetic circular birefringence of the metallic
magnetic nano-layers can lead to the extreme transmission directionality of
the layered structure.

\section*{Acknowledgements}

This research is supported by the Air Force Office of Scientific Research,
LRIR 09RY04COR.

\end{document}